\documentclass{article}
\usepackage[twocolumn]{emulateapj}
\usepackage{psfig}
\usepackage{apjfonts}

\def\farcs{\hbox{$.\!\!^{\prime\prime}$}}  
\def\lsim{\mathrel{\hbox{\rlap{\lower.55ex \hbox {$\sim$}}\kern-.0em
\raise.4ex \hbox{$<$}}}} 
\def\gsim{\mathrel{\hbox{\rlap{\lower.55ex \hbox {$\sim$}}\kern-.0em
\raise.4ex \hbox{$>$}}}} 
\newcommand{\gpm}[3]{$#1^{+#2}_{-#3}$}

\begin{document}

\title{Evidence for a supernova in reanalyzed optical and near-infrared
images of GRB\,970228} 

\author{T.J. Galama\altaffilmark{1},
N. Tanvir\altaffilmark{2},
P.M. Vreeswijk\altaffilmark{1},
R.A.M.J. Wijers\altaffilmark{3}, 
P.J. Groot\altaffilmark{1},
E. Rol\altaffilmark{1},
J. van Paradijs\altaffilmark{1,4},
C. Kouveliotou\altaffilmark{5,6},
A.S. Fruchter\altaffilmark{7},
N. Masetti\altaffilmark{8},
%A. Guarnieri\altaffilmark{8},
H. Pedersen\altaffilmark{9},
B. Margon\altaffilmark{10},
E.W. Deutsch\altaffilmark{10},
M. Metzger\altaffilmark{11},
%B. Soifer\altaffilmark{11},
%G. Neugebauer\altaffilmark{11},
L. Armus\altaffilmark{11},
S. Klose\altaffilmark{12},
B. Stecklum\altaffilmark{12}
}

\altaffiltext{1}{Astronomical Institute
`Anton Pannekoek', University of Amsterdam, \& Center for High Energy
Astrophysics, Kruislaan 403, 1098 SJ Amsterdam, The Netherlands}

\altaffiltext{2}{Department of Physical Sciences, University of
   Hertfordshire, College Lane, Hatfield, Herts, AL10 9AB, UK}

\altaffiltext{3}{Department of Physics and Astronomy, SUNY Stony
Brook, NY 11794-3800, USA} 

\altaffiltext{4}{Physics Department, University of Alabama in
Huntsville, Huntsville AL 35899, USA}

\altaffiltext{5}{Universities Space Research Asociation}

\altaffiltext{6}{NASA/MSFC, Code SD-40, Huntsville AL 35812, USA}

\altaffiltext{7}{Space Telescope Science Institute, 3700 San Martin
Drive, Baltimore, MD 21218, USA}

\altaffiltext{8}{CNR Bologna, Via P. Gobetti 101, 40129 Bologna, Italy}

\altaffiltext{9}{Copenhagen University Observatory, Juliane Maries Vej
30, D-2100, Copenhagen \"A, Denmark}

\altaffiltext{10}{Astronomy Dept., University of Washington, Box
351580, Seattle WA 98195-1580, USA}

\altaffiltext{11}{SIRTF Science Center, Caltech, MS 314-6 Pasadena CA  91125} 

\altaffiltext{12}{Th\"uringer Landessternwarte Tautenburg,
Sternwarte 5, D-07778, Tautenburg, Germany}

\begin{abstract}
We present B-, V-, R$_{\rm c}$-, I$_{\rm c}$-, J-, H-, K- and
K$^{'}$-band observations of the optical transient (OT) associated
with GRB\,970228, based on a reanalysis of previously used images and
unpublished data. In order to minimize calibration differences we have
collected and analyzed most of the photometry and consistently
determined the magnitude of the OT relative to a set of secondary
field stars. We confirm our earlier finding that the early decay of
the light curves (before March 6, 1997) was faster than that at
intermediate times (between March 6 and April 7, 1997).  At late times
the light curves resume a fast decay (after April 7, 1997).  The
early-time observations of GRB\,970228 are consistent with
relativistic blast-wave models but the intermediate- and late-time
observations are hard to understand in this framework. The
observations are well explained by an initial power law decay with
$\alpha$ = \gpm{-1.73}{0.09}{0.12} modified at later times by a
type-I$_{\rm c}$ supernova light curve.  Together with the evidence
for GRB\,980326 and GRB\,980425 this gives further support for the
idea that at least some GRBs are associated with a possibly rare type
of supernova.
\end{abstract}


%\notetoeditor{} 

\section{Introduction}
The $\gamma$-ray burst of February 28, 1997, was the first for which a
fading X-ray (Costa et al. 1997) and an optical counterpart (Groot et
al. 1997a; van Paradijs et al. 1997) were found.  After the
counterpart had weakened by several magnitudes, it was found to
coincide with an extended object (van Paradijs et al. 1997; Groot et
al. 1997b; Metzger et al. 1997a).  In subsequent observations with the
Hubble Space Telescope (HST) on March 26 and April 7, 1997, it was
found that the optical counterpart consisted of a point source and an
extended ($\sim 1^{\prime \prime}$) object (GAL), offset from the
point source by $\sim 0.5^{\prime \prime}$ (Sahu et al. 1997). The
extended object is most likely the host galaxy of GRB\,970228 and its
redshift was recently determined at $z = 0.695 \pm 0.002$ (Djorgovski
et al. 1999). The original light curve was compiled from data taken
from the literature (Galama et al. 1997).  Data obtained until April
1997 suggested a slowing down of the decay of the optical
brightness. However, the HST observations of September 1997 (Fruchter
et al. 1997, 1999; Castander and Lamb 1998a) showed that after April
1997 the light curve of the point source roughly continued a power-law
decay.

We have collected and reanalyzed the photometric information
(previously drawn from the literature and presented by Galama et
al. 1997), collected yet unpublished images and present here
reassessed Cousins V-, R$_{\rm c}$-, and I$_{\rm c}$-band optical
light curves and spectral flux distributions of the GRB counterpart
(Sect. \ref{sec:calib}, \ref{sec:rea}, \ref{sec:light} and
\ref{sec:spec}). We confirm the result of Galama et al. (1997) that
the early decay of the light curves (before March 6, 1997) was faster
than that at later times (between March 6 and April 7, 1997) and
discuss the observations in terms of the relativistic blast-wave and
blast-wave plus underlying supernova models in Sect. \ref{sec:dis}.
We find that the observations are well explained by an initial power
law decay with $\alpha$ = \gpm{-1.73}{0.09}{0.12} modified at later
times by a type-I$_{\rm c}$ supernova light curve. This is in
agreement with a recent suggestion by Reichart (1999). Together with
similar evidence for GRB\,980326 (Bloom et al. 1999) and GRB\,980425
(Galama et al. 1998) this gives further support for the idea that at
least some GRBs are associated with a possibly rare type of supernova.

\section{Reanalysis of previous observations}
\label{sec:rea}

We collected and analyzed most of the photometry of GRB\,970228.  In
Table \ref{Data1} we provide a log of all the photometry on
GRB\,970228, obtained in the B, V, R$_{\rm c}$, and I$_{\rm c}$
passbands (corresponding to the Johnson B and Cousins VRI system),
some photometry derived from unfiltered passbands and translated to B
or R$_{\rm c}$, the HST WFPC2 and STIS filters, and the near-infrared
J, H, K and K$^{'}$ passbands.  In order to minimize calibration
differences as a result of use of different filters/instruments and
differences in analysis we consistently determined the magnitude of
the optical transient (OT) relative to our calibration, which is
described in Sect. \ref{sec:calib}, using a set of secondary field
stars (Table \ref{Ref}).  In Sect. \ref{sec:details} and \ref{sec:gal}
we discuss each observation separately.

\subsection{Calibration}
\label{sec:calib}

As part of the ESO Very Large Telescope Unit 1 (VLT-UT1) Science
Verification (1998) a series of deep images of the GRB\,970228 field
were obtained in B, V and R$_{\rm c}$ (see Table \ref{Data1}). The
images were bias-subtracted and flat-fielded in the standard
fashion. Through the entire night observations of the Landolt fields
SA 99 (around star 253), PG1633+099, MARK A and TPHE (Landolt 1992)
were obtained allowing accurate photometric calibration for a number
of reference stars (see Table \ref{Ref}).  We determined the
transformation of the reference stars' magnitude to the Landolt system
by fitting for the zero point and the first order extinction and color
coefficients.

We also determined a calibration in B, V and R$_{\rm c}$ from
observations of the Landolt fields SA98 (around star 978), PG0918+029
and Rubin 149, taken as part of the Isaac Newton Telescope (INT)
observations of March 8 and March 9, 1997 (see Table \ref{Data1}).
These two calibrations agree, for stars C-H, to within 0.04, 0.02 and
0.10 magnitudes for B, V and R$_{\rm c}$, respectively. The difference
between the two calibrations is due to lesser depth of the INT
calibration and we estimated that the absolute calibration uncertainty
in each filter from the VLT-UT1 observations is 0.05 magnitudes.  We used
the VLT-UT1 magnitudes for the reference stars C-H and included stars A and
B from the INT calibration (stars A and B are not contained in the
field of the VLT-UT1 images), correcting for the small zero point offset
between the VLT-UT1 and INT calibration. The adopted reference stars'
magnitudes are listed in Table \ref{Ref}.

Observations of the field of GRB\,970228 in B, V, R$_{\rm c}$ and
I$_{\rm c}$ (for 240 sec each filter) and of Landolt field SA 98 were
obtained March 2, 1999 with the 1-m Johannes Kapteyn Telescope (JKT;
La Palma). Of stars A-H only star B was well detected, because of the
lesser depth of the JKT images. Therefore, to obtain the I$_{\rm
c}$-band calibration, we proceeded as follows. First, we determined
the magnitudes of star B and six more bright stars in the field of
GRB\,970228, correcting for the zero point, the atmospheric extinction
and also for a first-order color term. We then used the I$_{\rm
c}$-band calibration of these bright field stars to calibrate stars
A-H using the William Herschel Telescope (WHT) March 8, 1997 I$_{\rm
c}$-band image; these magnitudes are listed in Table \ref{Ref}. For
consistency: the calibration in B, V, and R$_{\rm c}$ for star B is in
good agreement with the calibration determined from the VLT-UT1 and
INT data (better than 0.07 mag).

Near-infrared images in the J and K band of the field of GRB\,970228
and of standard star FS 14 (Casali and Hawarden 1992) were obtained on
March 30.2, 1997 UT with the Near Infrared Camera (NIRC) at the Keck-I
10-m telescope. From these observations we obtained a calibration of
the nearby star (star H); we found $J = 20.27 \pm 0.10$ and $K = 19.67
\pm 0.10$. Another set of near-infrared images of the field of
GRB\,970228 and of standard star HD 84800 (Casali and Hawarden 1992)
were obtained in J, H and K$^{'}$ with the Calar Alto 3.5-m telescope on
March 17.8, 1997 UT. From these observations we found for the nearby
star (star H): $J = 20.20 \pm 0.15$, $H = 19.16 \pm 0.15$ and $K^{'} =
19.59 \pm 0.28$. The two calibrations are in good agreement with each
other and we adopted their weighted averages: $J = 20.25 \pm 0.10, H =
19.16 \pm 0.15$ and $K = 19.66 \pm 0.10$.	

\subsection{The details of each observation}
\label{sec:details}

The earliest image of the OT was reported by Pedichini et
al. (1997). The reported magnitude was given relative to the nearby
late-type star (LTS; star H in Table \ref{Ref}). This observation was
obtained without a filter; we have transformed this unfiltered
magnitude to the R$_{\rm c}$ band, using the reported filter
characteristics (Pedichini et al. 1997), taking $V-I_{\rm c}$ = 0.50
$\pm$ 0.23 (from the February 28.99, 1997 UT WHT observations) and
assuming a power-law spectrum $F_{\nu} \propto \nu^{\beta}$. The
resulting value $R_{\rm c} = 20.0 \pm 0.5$ is consistent with that
reported by Pedichini et al. (1997) and Galama et al. (1997).

In the images of Guarnieri et al. (1997)\footnote{Note that in Galama
et al. 1997 the time of the observation of Guarnieri et al. 1997 is
given incorrectly.}, the OT was blended with the nearby late-type
star. We analyzed their February 28.83, 1997 UT R$_{\rm c}$-band image
and used a large aperture to include both the OT and the late-type
star. The resulting magnitude $R_{\rm c}$ = 20.30 $\pm$ 0.28 (OT + GAL
+ LTS) in our calibration is about 0.8 mag brighter than that reported
by Guarnieri et al. (1997); correcting for the LTS we found $R_{\rm c}$
= 20.58 $\pm$ 0.28 (OT + GAL). We have not included the subsequent
measurements and upper-limits as given by Guarnieri et al. (1997) as
they are consistent with detections of the late-type star only.

Reanalysis of our WHT, INT and New Technology Telescope (NTT; La
Silla) images have led to marginal corrections; all corrections, apart
from the V band February 28, 1997 (which is 0.3 mag off), are within
the errors as presented in Galama et al. (1997).  We also present an
unpublished R$_{\rm c}$-band non-detection (INT March 8.89, 1997 UT).

The March 3.1, 1997 UT observation by Margon et al. (1997) with the
Astrophysical Research Consortium's 3.5-m telescope at the Apache
Point Observatory was performed using an unfiltered (UF) 
passband (blue). As the OT was faint in the APO image we determined its
position relative to a number of nearby field stars in the February 28.99,
1997 UT V-band WHT image (where the OT is bright) and performed a
rotation plus translation to obtain the corresponding pixel position
in the APO image; subsequently we have used this position to center the
aperture to determine the OT's magnitude (the aperture was chosen such
that the extended emission was also contained). The unfiltered OT's
magnitude was translated to the B passband by fitting an empirical
linear relation between the $UF-B$ magnitude difference and the $B-V$
magnitude difference determined from the reference stars A-H (Table
\ref{Ref}). We found $UF-B$ = 24.03(65) + 0.45(11)($B-V$)
($\chi^{2}$ = 1.2 with 3 d.o.f). We obtained a rough estimate of the
color of the OT by fitting a temporal power law decay, $F_{\nu}
\propto t^{-\alpha}$, to the V-band data, interpolating the fit to
March 9.85, 1997 UT and using the B-band measurement at that same
epoch; we found $B-V$ = 1.10 $\pm$ 0.34 on March 9.85, 1997
UT. Assuming that the color of the OT did not substantially change
this corresponds to $B$ = 24.5 $\pm$ 0.7 on March 3.1, 1997
UT. (Note that the error is not dominated by the assumed color). This
is about 1.4 $\sigma$ different from the original value (Margon et
al. 1997).

Apart from the March 4.9, 1997 UT V-band image (presented by Galama et
al. 1997) additional, and yet unpublished, unfiltered (red) and B-band
images were obtained with the Nordic Optical Telescope (NOT).  The OT
was clearly detected in the unfiltered March 4.9, 1997 UT image but
not in the V and B-band images of March 4.9, 1997 UT nor in the
unfiltered image of March 3.9, 1997 UT.  Similar to the procedure for
the APO data we determined the pixel position of the OT from the
February 28.99 V-band WHT image to obtain the magnitude of the OT in
the unfiltered NOT March 4.9, 1997 UT image. The unfiltered OT's
magnitude was translated to the R$_{\rm c}$ passband by fitting an
empirical linear relation between $UF-R_{\rm c}$ and $V-R_{\rm c}$
determined from the reference stars A-I (Table \ref{Ref}). We found
$R_{\rm c}$ = 22.97(26) + 0.48(16)($V-R_{\rm c}$) ($\chi^{2}$ = 4.3
with 6 d.o.f). Taking $V-R_{\rm c}$ = 1.03 (corresponding to $V-I_{\rm
c}$ = 2.24 $\pm$ 0.14, as observed for the OT with the HST March 26
and April 7, 1997; see below) we found $R_{\rm c}$ = 23.46 $\pm$
0.32. In a similar way, for $V-I_{\rm c}$ = 0.50 $\pm$ 0.23 (WHT February
28.99, 1997) we obtained $R_{\rm c}$ = 23.10 $\pm$ 0.27; we adopted
the mean value $R_{\rm c}$ = 23.28 $\pm$ 0.40.  Using the same
empirical relation between $UF-R_{\rm c}$ and $V-R_{\rm c}$ we obtained
$R_{\rm c} > 22.7 \pm 0.2$ for the unfiltered March 3.9, 1997 UT image
(using a conservative $V-I_{\rm c}$ = 2.24).

We also determined the pixel position of the OT to obtain its
magnitude in the Keck images of March 6.3 and April 5.8, 1997 UT,
using again the February 28.99, 1997 UT V-band WHT image. We found the
March 6.3 R$_{\rm c}$ magnitude, relative to our calibration, to be
0.4 mag brighter than that reported by Metzger et al. (1997a); the
April 5.8 magnitude is consistent with that reported by Metzger et
al. (1997b).

We have not analyzed the Palomar 200$^{''}$ September 4, 1997
observation of Djorgovski et al. (1997); it is consistent with
detection of the nebulosity only.

We reanalyzed the GRB\,970228 HST WFPC2 (F606W and F814W) and STIS
images of Fruchter et al. (1999). For the details of the observations
we refer the reader to that work. We determined the WFPC2 magnitudes
of the OT and stars C, G and H, using a 2 pixel aperture radius. We
corrected the instrumental magnitudes for: (i) the `long versus short
anomaly' (Casertano and Mutchler 1998; corrections are --0.16, --0.24,
--0.12 and --0.19 mag for the WFPC2 observations of March 26 (V),
April 7 (V), March 26 (I), April 7 (I), respectively), (ii) the effect
of non-optimal Charge Transfer Efficiency (CTE; Whitmore and Heyer,
1997; Whitmore 1998; corrections are $\sim$ --0.06 mag), and (iii) the
geometric distortion of WFPC2 images (Holtzman et al. 1995a;
corrections are smaller than 0.03 mag). Next we determined the
aperture correction from 2 to 11 pixels, used the instrumental zero
points from Baggett et al. (1997) and the color corrections from
Holtzman et al. (1995b) to obtain the final V- and I$_{\rm c}$-band
magnitudes. The errors were determined by adding in quadrature the
measurement error (Poisson noise), the errors in the above-mentioned
corrections, the aperture correction error and a 2\% uncertainty for
the instrumental zero points. For the OT the error in the magnitude is
dominated by the `long versus short anomaly' correction (the
correction is larger for fainter objects). The results for the OT are
given in Table \ref{Data1}. We found systematically brighter
magnitudes than Fruchter et al. (1999) and Castander and Lamb (1998a),
which can be explained by the `long versus short anomaly' correction
that we applied. We found the OT to be redder ($V-I_{\rm c}$ = 2.22
$\pm$ 0.25 and $V-I_{\rm c}$ = 2.26 $\pm$ 0.30, March 26 and April 7,
1997 UT, respectively) than found by Fruchter et al. (1999), but
consistent with the result of Castander and Lamb (1998a).  For stars
C, G and H we obtained ($V$, $I_{\rm c}$) = (21.76 $\pm$ 0.05, 20.10 $\pm$
0.05), (23.46 $\pm$ 0.05, 22.52 $\pm$ 0.07) and (22.85 $\pm$ 0.04,
21.04 $\pm$ 0.04), respectively. Comparing this with the ground-based
calibration (see Table \ref{Ref}) we found the agreement for the bright
star C to be excellent (better than 0.02 mag), for star H we found 0.15
and 0.12 mag differences and the V-band HST magnitude of star G was
0.22 mag off.

The HST STIS observations were taken in unfiltered mode, meaning that
the response covers a wide wavelength range from about 2000 to 10000
\AA.  To convert the observed counts into broadband magnitudes we had
to assume therefore something about the spectral energy distribution.
Our procedure was as follows: (i) measure the counts in a 2-pixel
radius aperture; (ii) add an aperture correction of -1.15 mags based
on brighter stars in the field, assuming that 10\% of the light falls
outside a large 11 pixel radius aperture; (iii) assume a power law
slope with spectral index chosen to be consistent with the earlier
WFPC2 observations ($\beta = -4.7$), and use the SYNPHOT routine
CALCPHOT to determine the corresponding broad-band magnitudes to
explain the observed counts. The result, corrected for the small 12\%
renormalization found by Landsman (1997), was $V = 28.80$, $R_{\rm c}
= 27.69$ and $I_{\rm c} = 26.54$.  Although the uncertainty in the
measured counts is only 10\%, the considerable dependence on the shape
of the assumed spectrum, and the difficulty of absolutely calibrating
the wide STIS passband (see below), leads us to adopt an uncertainty
of 0.4 magnitudes for each band. 

To check the STIS result we tried an alternative calibration using our
WFPC2 photometry for the two brightest field stars, but still relying
on SYNPHOT to calculate the appropriate colour correction, and found a
STIS magnitude of $V = 28.58$. This discrepancy is larger than the
formal errors, and is probably indicative of the difficulty of
precisely calibrating the very wide unfiltered STIS CCD passband.

We collected and rereduced the near-infrared J- and K-band images of
Soifer et al. (1997), obtained on March 30.2 and 31.2, 1997 UT with
NIRC on the Keck-I 10-m telescope, and the J-, H- and K$^{'}$-band
images of Klose, Stecklum and Tuffs (1997), obtained with the Calar
Alto 3.5-m telescope on March 17.8, 1997 UT using the near-infrared
camera MAGIC.  The infrared frames were reduced by first removing bad
pixels and combining at least five frames around each object image to
obtain a sky image. This sky image was then subtracted after scaling
it to the object image level.

A source was detected in both bands J and K in the Keck-I
observations. Relative to the nearby star (star H) we found $J = 23.27
\pm 0.15$ (OT + GAL; sum of the March 30.3 and March 31.2, 1997 UT
data) and $K = 22.85 \pm 0.25$ (March 30.2, 1997 UT) within a large
0\farcs75 aperture radius. The J-band magnitude is consistent with the
earlier determination (Soifer et al. 1997), but our K-band
determination is 0.8 mag fainter. Fruchter et al. (1999) detected the
host galaxy at $K = 22.8 \pm 0.3$ and so our result is consistent with
detection of the host galaxy only.

We did not detect the source in the Calar Alto J, H and K$^{'}$ images and
determined the following limiting magnitudes: $J > 21.45$, $H > 20.02$
and $K^{'} > 20.13$ (3 $\sigma$). We confirm the non detections in H and
K$^{'}$ by Klose et al. (1997), but do not confirm their detection of the
source in J\footnote{The J-band magnitude reported by Klose et al.
(1997) is affected by a transient warm pixel at the position of the
source that was not obvious in the bad pixel mask.}.

\subsection{The host galaxy}
\label{sec:gal}

The host galaxy of GRB\,970228 is clearly detected in the HST WFPC2
images, but at low signal-to-noise.  Working with cosmic-ray and
hot-pixel cleaned images, we first subtracted point-spread functions
positioned and scaled to remove the image of the OT from each.
For the two F606W images we then performed aperture photometry in
apertures of increasing size, taking a sky estimate from the
surrounding area which was free of other objects.  We found that the
counts reached an asymptote, within the noise, by a radius of 25
pixels ($\sim$ 1\farcs1).  We calibrated this using the procedure of
Holtzman (1995b), finding an average $V_{606} = 25.5 \pm 0.3$, where
the error is dominated by the photon and read noise.

In the F814W image the host galaxy is even fainter, and we estimated
its color by comparing the counts within a 7 pixel radius aperture
with the counts in the same aperture on the F606W image.  Thus we
found a color $V_{606} - I_{814} \sim 1.1$ which is, correcting for
the Galactic extinction, reasonable for a galaxy at this magnitude
(Smail et al. 1995).  This could then be used to estimate color
corrections to the standard systems, and we concluded $V = 25.75 \pm
0.35$ and $I_{\rm c} = 24.7 \pm 0.5$.

The host galaxy is much better detected on the STIS image, although we
must assume a spectral shape in order to estimate broad-band
magnitudes.  Using CALCPHOT we found a good match to the WFPC2 colours
with a suitably redshifted and reddened S$_{\rm c}$ galaxy
spectrum\footnote{From the Kinney-Calzetti atlas, see the Synphot
User Guide 1998, Space Telescope Science Institute.}.  Based on this,
we infer $V = 25.77$, $R_{\rm c} = 25.22$ and $I_{\rm c} = 24.73$ from
the STIS data, which is in good agreement with the other
determinations.  We adopt an error of 0.2 mag to account for the
systematic uncertainty introduced by the necessity of assuming the
spectral shape.

We did not detect the host galaxy of GRB\,970228 in the ESO VLT-UT1
Science Verification images and determined the 3-$\sigma$ limit for
detection using an aperture radius of 1.5 times the seeing, which
corresponds to $\sim$ 1\farcs5. The results are shown in Table
\ref{Data1}. The V and R$_{\rm c}$-band upper limits for the host
galaxy show that the host galaxy is fainter than the earlier
determination (Sahu et al. 1997), but consistent with the result of
Fruchter et al. (1999), Castander and Lamb (1998a) and with our
estimate from the WFPC2 images.

\section{The light curves}
\label{sec:light}

In the interpolations to the R$_{\rm c}$ band (see
Fig. \ref{Lightcurve}) we have assumed that the spectra of both the
point source and the host galaxy, are smooth (i.e., not dominated by
emission lines). We have used the relation between the color indices
$V-R_{\rm c}$ and $V-I_{\rm c}$ given by Th\'e et al. (1984) for
late-type stars; for bluer stars we have inferred this relation from
the tables given by Johnson (1966) for main-sequence stars and the
color transformations to the Cousins VRI system given by Bessel
(1976).  We have tested the validity of these color-color relations
from numerical integrations of power law flux distributions and of
Planck functions, and conclude that if the flux distributions of the
OT and the host galaxy are smooth, the uncertainty in the interpolated
R$_{\rm c}$ magnitude is unlikely to exceed 0.1 mag (Galama et
al. 1997). For the HST September 4, 1997 observation we have assumed
that the colors of the OT remained constant during the late-time decay
(i.e., taking the observed $V-I_{\rm c}$ = 2.24 from the HST March 26
and April 7, 1997 observations; see Sect. \ref{sec:details}).  Finally, we
corrected for the contribution of the host galaxy emission ($V$ = 25.77
$\pm$ 0.20, $R_{\rm c} = 25.22 \pm 0.20$ and $I_{\rm c} = 24.73 \pm 0.20$)
and obtained the V-, R$_{\rm c}$-, and I$_{\rm c}$-band light curves of
the OT (Fig. \ref{Lightcurve}).
 
We have fitted power laws, $F \propto t^{\alpha}$, to the light curves
(ignoring upper limits) and found: $F_{{\rm V}} = (11.7 \pm
1.0)\,t_{\rm d}^{-1.32 \pm 0.04} \mu$Jy ($\chi^2$ = 0.2 with 2 d.o.f),
$F_{{\rm R}_{\rm c}} = $\gpm{12.7}{1.2}{1.8}$\,t_{\rm d}^{-1.14 \pm
0.04} \mu$Jy ($\chi^2$ = 9.7 with 8 d.o.f), and $F_{{\rm I}_{\rm c}} =
$\gpm{15.2}{3.1}{2.5}$\,t_{\rm d}^{-0.92 \pm 0.06} \mu$Jy ($\chi^2$ =
4.5 with 2 d.o.f), where $t_{\rm d}$ is the time in days. These fits
are indicated with dotted lines in Fig. \ref{Lightcurve}. We also
fitted a power law to the R$_{\rm c}$-band light curve at intermediate
times (between 3 and 50 days) and found $F_{{\rm R}_{\rm c}} =
$\gpm{3.4}{2.4}{1.4}$\,t_{\rm d}^{-0.73 \pm 0.17} \mu$Jy ($\chi^2$ =
1.3 with 4 d.o.f).  The fit is indicated with a thick line in
Fig. \ref{Lightcurve}.  For the late-time detections (after 20 days)
we found: $F_{{\rm V}} = (11.7 \pm 1.3)\,t_{\rm d}^{-1.14 \pm 0.21}
\mu$Jy ($\chi^2$ = 0.2 with 1 d.o.f), $F_{{\rm R}_{\rm c}} =
$\gpm{21.5}{23}{11}$\,t_{\rm d}^{-1.27 \pm 0.20} \mu$Jy ($\chi^2$ =
0.3 with 1 d.o.f) and $F_{{\rm I}_{\rm c}} =
$\gpm{50}{48}{24}$\,t_{\rm d}^{-1.25 \pm 0.19} \mu$Jy ($\chi^2$ = 0.9
with 1 d.o.f), where $t_{\rm d}$ is the time in days. These fits are
indicated with dashed lines in Fig. \ref{Lightcurve}.

\section{Optical/near-infrared to X-ray spectra}
\label{sec:spec}

For three epochs we have reconstructed the spectral flux distribution
of the OT, corresponding to: (i) 1997 February 28.99 UT (V, R$_{\rm c}$ and
I$_{\rm c}$ data from Table \ref{Data1} and 2-10 keV flux from Costa
et al. 1997), (ii) March 9.89 UT (R$_{\rm c}$ band measurement from Table
\ref{Data1} and ROSAT 0.1-2.4 keV unabsorbed flux from Frontera et
al. 1998), and (iii) March 30.8 UT (See Fig. \ref{Spectra}). We subtracted
the host galaxy flux from ground-based measurements, 
corrected the OT fluxes for Galactic foreground absorption, A$_V$ =
0.78 $\pm$ 0.12 (from Schlegel, Finkbeiner and Davis 1998; but see
also the discussion by Gonz\'alez, Fruchter, and Dirsch 1999, and
Castander and Lamb 1998b), and brought measurements to the same epoch
using the power-law fits of Sect. \ref{sec:light}. We note that the
spectral flux distribution on March 30.8 UT (Fig. \ref{Spectra}) is
different from that of Reichart (1999) because in our reanalysis the
optical transient was not detected in the K band. The resulting
spectral flux distribution therefore supersedes that of Reichart
(1999) who obtained the K-band magnitudes of the OT and the host from
the literature.

We fitted the optical to X-ray spectral flux distribution of the first
epoch (February 28.99, 1997 UT) with a power law and an exponential
optical extinction law, $F_\nu \propto \nu^{\beta}e^{-\tau}$, where we
assumed that the extinction optical depth, $\tau \propto \nu$. The
data did not allow determination of the host galaxy extinction; the
fit provided a negative extinction (A$_V < 0.50$; 90 \% confidence)
and was subsequently fixed at zero.  We found a spectral slope, $\beta
= -0.780 \pm 0.022$ (reduced $\chi$-squared, $\chi^{2}_{\rm r}$ = 4/2).
For comparison, the X-ray power-law photon index observed $\sim$ 12
hours after the burst (2-10 keV; Costa et al. 1997) corresponds to a
spectral slope $\beta = -1.1 \pm 0.3$.

\section{Discussion}
\label{sec:dis}

\begin{figure}[b!] % fig 1
\centerline{\psfig{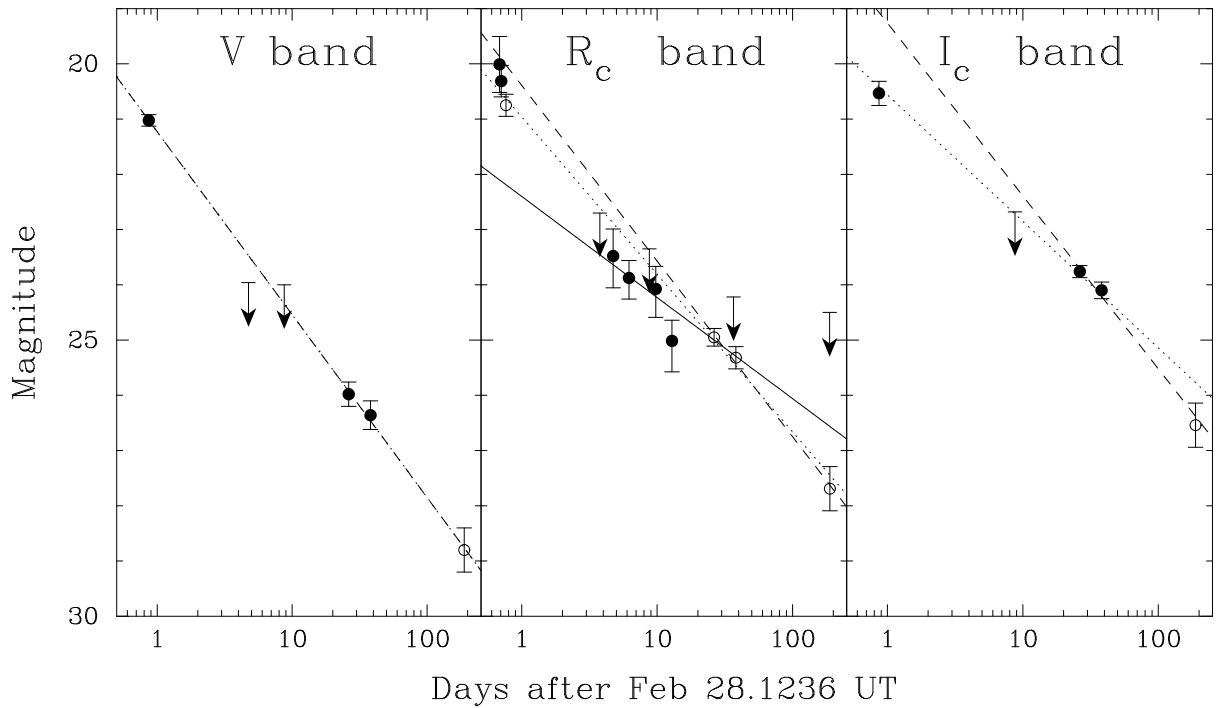}}
\caption{The V-, R$_{\rm c}$- and I$_{\rm c}$-band lightcurves of
GRB\,970228 (magnitude versus time). R$_{\rm c}$-band data obtained
from interpolations between V and I$_{\rm c}$ are indicated with open
symbols (see the discussion for details).  Indicated are power-law
fits, $F_{\nu} \propto t^{\alpha}$, to the detections (ignoring upper
limits). Fits to all points are indicated by the dotted lines and fits
to the late-time lightcurve (after 20 days) are indicated by the
dashed lines (for the V band they overlap). Also indicated is a fit to
the R$_{\rm c}$-band light curve at intermediate times (between 3 and
50 days; thick line).}
\label{Lightcurve}
\end{figure}

Most X-ray and optical/infrared afterglows display a power law
temporal decay; this is a prediction for relativistic blast-wave
models of GRB afterglows, in which a relativistically expanding shock
front, caused by an energetic explosion in a central compact region,
sweeps up the surrounding medium and accelerates electrons in a strong
synchrotron emitting shock (e.g., M\'eszar\'os and Rees 1997).
Assuming a single power law and ignoring upper limits for the moment,
the V-, R$_{\rm c}$- and I$_{\rm c}$-band light curves
(Fig. \ref{Lightcurve}) indicate that the power law decay rate is an
increasing function of frequency; for the V, R$_{\rm c}$ and I$_{\rm
c}$ band the power-law decay index $\alpha$ = $-1.32 \pm 0.04$, $-1.14
\pm 0.04$, and $-0.92 \pm 0.06$, respectively. This implies that as a
function of time the OT becomes redder and redder. Such a relation has
not been observed previously, and is not what is expected in standard
relativistic blast-wave models, where the power law index is, to first
order, independent of frequency.

Taking the upper limits into account, we find that the March 4.87,
1997 UT NOT V-band non-detection is significantly below the fit. Also,
the R$_{\rm c}$-band light curve suggests that the optical emission is
weaker around that time (between 3 and 10 days) than expected from a
single power law. The V and R$_{\rm c}$ data suggest that the initial
decline was faster than the decline at intermediate times. From the
V-band February 28.99 and March 4.87, 1997 UT data we find that the
early-time decay was steeper than $\alpha = -1.4$ ($3\sigma$), while
at intermediate times (around $\sim$ 10 days) it was $\alpha = -0.73
\pm 0.17$ (R$_{\rm c}$).  This structure in the light curve was
recognized early on by Galama et al. (1997), who showed that the early
decay of the light curves (before March 6, 1997) was faster than that
at later times (between March 6 and April 7, 1997). The earlier
analysis of the March 26 and April 7, 1997 HST WFPC2 images yielded
brighter estimates of the host galaxy magnitudes (Sahu et al. 1997)
than our reanalysis (see also Fruchter et al. 1999; Castander and Lamb
1998a). Because of this, the correction to the ground-based data for
the host galaxy emission applied in this work is smaller than that of
Galama et al. (1997), and thereby the effect is somewhat reduced, but
it is still present (see also Reichart 1999) and consistent with the
earlier result.

The late-time light curves show that the rate of decay increased again
after $\sim$ 35 days. The late-time decays in V, R$_{\rm c}$ and
I$_{\rm c}$ are consistent with being frequency independent (i.e.,
similar decays in V, R$_{\rm c}$ and I$_{\rm c}$), with a power-law
index $\alpha = -1.2$. The back extrapolation of the late-time fit in
the V-band predicts quite well the earliest V-band data, but this is
not the case for the R$_{\rm c}$- and I$_{\rm c}$-band light curves.
We conclude that a single power-law is not a good representation of
the light curves; the overall shape of the lightcurves is that of an
initial fast decay (before $\sim$ March 6, 1997) followed by a
`plateaux' (slow decay; between $\sim$ March 6 and $\sim$ April 7,
1997) and finally the light curves resume a fast decay (after $\sim$
April 7, 1997).

For synchrotron radiation, by a power-law energy distribution of
electrons, at low frequencies, and for an adiabatic evolution of the
blast wave we can distinguish the following two cases, each of which
has its own relation between the spectral slope $\beta$ and the power
law decay index, $\alpha$ (Sari, Piran and Narayan 1998): (i) both the
peak frequency $\nu_{\rm m}$ and the cooling frequency $\nu_{\rm c}$
are below the optical waveband.  Then $\alpha = 3\beta/2 + 1/2 = -0.67
\pm 0.03$ (taking the measured spectral slope, $\beta = -0.78$ on
February 28.99, 1997 UT) (ii) $\nu_{\rm m}$ has passed the optical
waveband, but $\nu_{\rm c}$ has not yet. In that case $\alpha$ =
3$\beta/2 = -1.17 \pm$ 0.03. The temporal decay values measured at
early times are in reasonable agreement with case (ii), but certainly
not with case (i), suggesting that the peak frequency $\nu_{\rm m}$ is
below the optical passband and that the cooling frequency $\nu_{\rm
c}$ is near or above the X-ray (2-10 keV) passband; such a high
cooling frequency appears to be the fairly general case in GRB
afterglows (e.g., Bloom et al. 1998; Vreeswijk et al. 1999; Galama et
al. 1999).  The fit to the optical to X-ray spectrum of February
28.99, 1997 UT suggests that extinction at the host is not very
significant ($A_V < 0.5$; corresponding to a restframe extinction $A_V
< 0.3$). At the second epoch we measured a spectral slope, $\beta =
-0.52 \pm 0.08$, on March 9.89, 1997 UT.  For case (ii) this would
correspond to a power-law decay of $\alpha = -0.78 \pm 0.12$,
consistent with a shallower decay at intermediate times. We thus find that the
observations of the early-time afterglow are roughly consistent with
the predictions of simple relativistic blast-wave models (see also
Wijers, Rees and {M\'esz\'aros} 1997; Waxman 1997; Reichart 1997),
except that we find evidence for a flattening of the decay rate around
$\sim$ 10 days.

\begin{figure}[b!] % fig 2
\centerline{\psfig{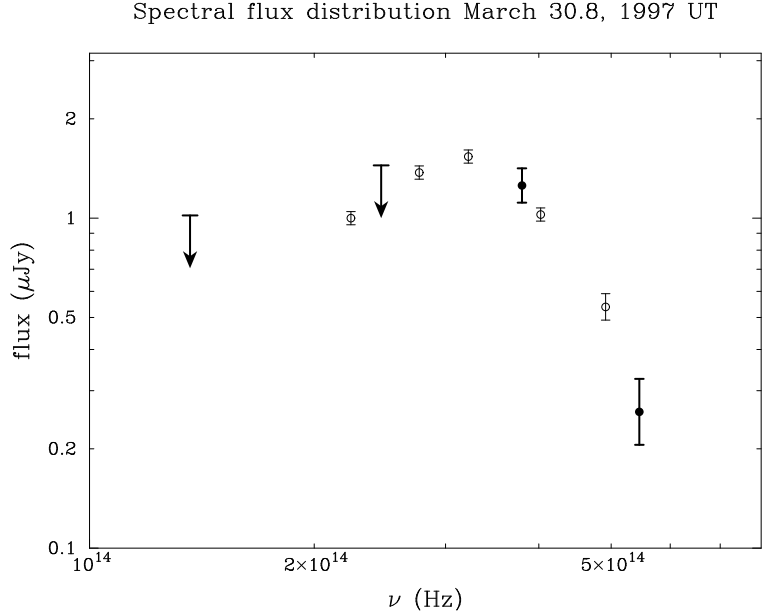}}
\caption{The optical/near-infrared spectral flux distribution
of GRB\,970228 at March 30.8, 1997 UT ($\bullet$ and upper-limit
arrow); from left to right: K-band detection (Table \ref{Data1})
translated into an upper limit because it is consistent with detection
of the host galaxy only (detection magnitude minus 3 $\sigma$), J-band
detection (Table \ref{Data1}) translated into an upper limit due to
unknown J-band host magnitude (detection magnitude minus 3 $\sigma$),
V and I$_{\rm c}$ HST measurements from Table \ref{Data1}. Also shown
is the spectral flux distribution of SN\,1998bw redshifted to the
redshift of GRB\,970228 ($\circ$).  Note that the spectral flux
distribution is different from that of Reichart (1999) (See
the discussion in Sect. \ref{sec:spec}).}
\label{Spectra}
\end{figure}

As mentioned above, the frequency dependence of the rate of decay of the
light curves corresponds to a reddening of the optical afterglow with time
($V-I_{\rm c} = 0.50 \pm 0.23$ at February 28.99, 1997 UT and $V-I_{\rm c}
= 2.24 \pm 0.14$ at March 26 and April 7, 1997). We are confident
about this large change in color as the HST calibration of three field
stars agrees well with that of the ground-based calibration. Assuming
a power-law spectrum, the optical spectral slopes (Galactic extinction
corrected) from the intermediate-time V and I$_{\rm c}$ HST observations are
$\beta = -3.8 \pm 0.7$ (March 26, 1997) and $\beta = -4.0 \pm 0.8$
(April 7, 1997). These steep spectra cannot be explained by extinction
at the host galaxy; we find a staggering A$_V$ = 4.5 $\pm$ 1.0
(fitting as before for an exponential optical extinction law, $F_\nu
\propto \nu^{\beta}e^{-\tau}$, with $\tau \propto \nu$, and fixing the
spectral slope at $\beta = -0.78$), which would imply that the extinction
increased with time from negligible (A$_V <$ 0.5) to very significant
(A$_V =$ 4.5 $\pm$ 1.0); this seems quite unlikely. A steepening of
the spectral slope may be expected from a number of spectral
transitions: e.g, the passage of the cooling frequency or the
transition to a Sedov-Taylor non-relativistic expansion (e.g., Sari et
al. 1998; Waxman, Kulkarni and Frail 1998). However, such spectral
transitions will be accompanied with an increase in the rate of decay
of the light, which we do not observe here.  The discrepancy is even
larger if one considers the spectral flux distribution around the
third epoch (March 30.8, 1997 UT; Fig. \ref{Spectra}); the spectrum
shows an unexpected turnover around $3 \times 10^{14}$ Hz (Fruchter et
al. 1999). This break plus the very red spectrum in V and I$_{\rm c}$
in the intermediate-time optical flux distribution is not consistent with the
predictions of simple relativistic blast-wave models (see also
Reichart 1999; Fruchter et al. 1999).

What then could explain these observations?  Bloom et al. (1999)
observed a similarly red spectrum for the late-time emission from
GRB\,980326. The authors argue that this is due to a supernova (SN)
that dominated the light at late times. Reichart (1999) suggests that
this is also the case for GRB\,970228. In this hypothesis the early
light curve is dominated by a (power-law decaying) GRB afterglow
(produced by the relativistic blast wave), while at late times the
light curve contains a significant contribution from an underlying
supernova. Such would be the natural outcome of the
`collapsar/microquasar' models in which GRBs arise in jets that are
formed in the core collapse to a black hole of massive stars (Woosley
1993; Paczy\'nski 1998). Such massive stars will, prior to collapse,
have lost their hydrogen envelopes and so we expect the supernova to
be of type I$_{\rm b/c}$ (Woosley 1993; Woosley, Eastman and Schmidt
1999; Bloom et al. 1999).

\begin{figure}[b!] % fig 3
\centerline{\psfig{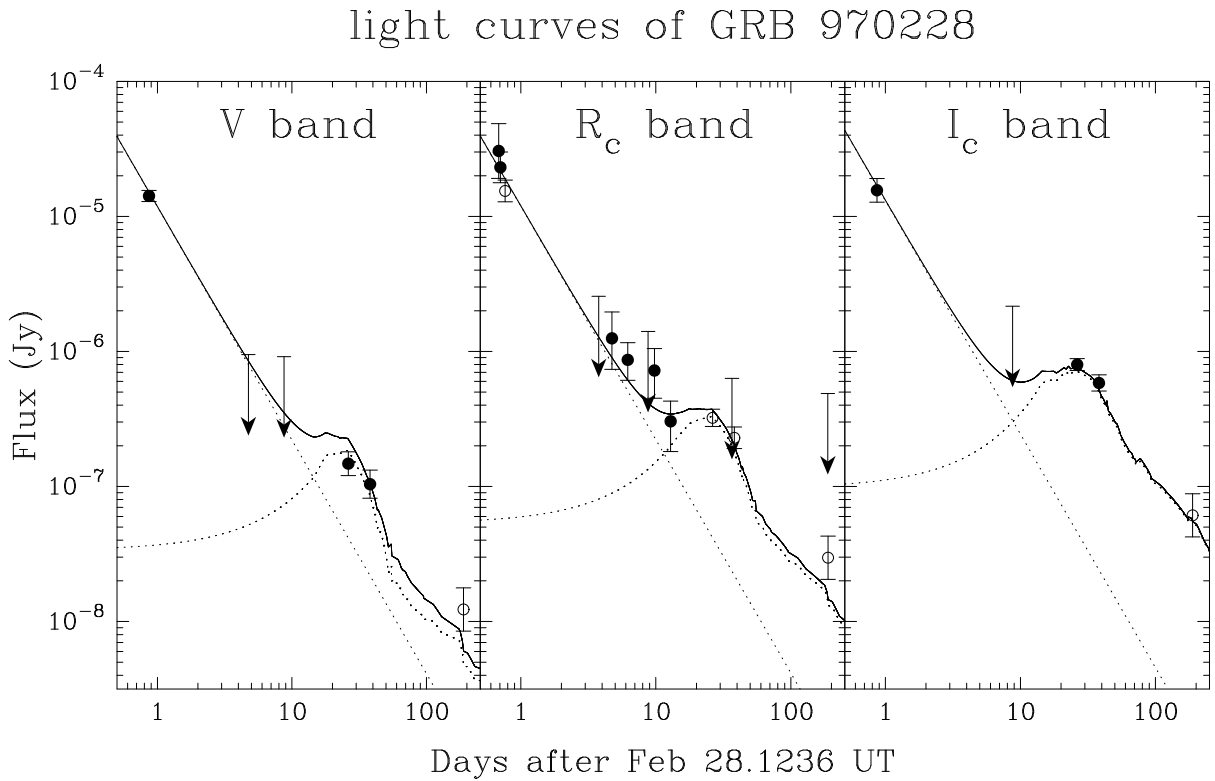}}
\caption{The V-, R$_{\rm c}$-, and I$_{\rm c}$-band lightcurves of
GRB\,970228 (fluxes versus time). The dotted curves indicate power-law
decays with $\alpha = -1.73$, and redshifted SN\,1998bw light
curves. The thick line is the resulting sum of SN and power-law decay
light curves.}
\label{Lightcurve2}
\end{figure}

Following Bloom et al. (1999) and Reichart (1999) we constructed
simulated supernova light curves and spectral flux distributions as
follows.  First we combined three sets of Galactic extinction
corrected U-, B-, V-, R$_{\rm c}$- and I$_{\rm c}$-band observations
of the well-observed, unusual type-I$_{\rm c}$ SN\,1998bw (Galama et
al. 1998; McKenzie and Schaefer 1999; Vreeswijk et al. 1999), which is
likely to be associated with GRB\,980425 (Galama et al. 1998; Kulkarni
et al. 1998)\footnote{The association between SN\,1998bw and
GRB\,980425 has been considered uncertain because the behavior of a
another source in the error box of GRB\,980425, the fading X-ray
source 1SAX J1935.3-5252, resembled that of a typical GRB
afterglow. However, reanalysis of NFI BeppoSAX observations of the
GRB\,980425 field by Pian et al. (1999) shows that its behavior is
unlike that of previously observed GRB afterglows. SN\,1998bw remains
not typical for GRB afterglows because of its low redshift, $z =
0.0085$.}. We then redshifted the lightcurves to the redshift of
GRB\,970228, $z = 0.695$ (Djorgovski et al. 1999), which consists of
wavelength shifting, time-profile stretching, and dimming (we assumed
a cosmology with $H_0 = 70$ km s$^{-1}$ Mpc$^{-1}$ and $\Omega_0 =
0.3$). To obtain the corresponding redshifted light curves in the V,
R$_{\rm c}$ and I$_{\rm c}$ bands we assumed a power-law spectral
shape to interpolate (and extrapolate) the redshifted SN\,1998bw light
curves to the central wavelengths of the V, R$_{\rm c}$ and I$_{\rm
c}$ bands. (Note that the redshifted central wavelength of the U band
is near V and R$_{\rm c}$, and for the B band is near I$_{\rm c}$.)
Finally we corrected the SN light curves for the Galactic extinction
in the direction of GRB\,970228 ($A_V = 0.78 \pm 0.12$).  These light
curves are shown in Fig. \ref{Lightcurve2}.

The supernova spectral flux distribution corresponding to the epoch of
March 30.8, 1997 UT (30.7 days after GRB\,970228) is shown in
Fig. \ref{Spectra} (here we did the opposite: we corrected the
GRB\,970228 spectral flux distribution for Galactic extinction). The
resemblance of the GRB\,970228 and the redshifted SN\,1998bw spectral
flux distributions is remarkable (see also Reichart 1999).

At early times the R$_{\rm c}$-band light curve (see
Fig. \ref{Lightcurve2}) will not be very much dominated by SN
emission. If we attribute the early part ($t < 8$ days) of the R$_{\rm
c}$-band light curve to GRB afterglow and fit a power-law temporal
decay to the data we find a temporal decay index $\alpha = -1.46 \pm
0.16$ ($\chi^{2}$ = 1.5 with 3 d.o.f). This value of the decay
constant is consistent with that in X rays, $\alpha$ =
\gpm{-1.33}{0.11}{0.13} during the first $\sim$ 4 days (2-10 keV;
Costa et al. 1997) and $\alpha$ = \gpm{-1.50}{0.35}{0.23} up to $\sim$
10 days (0.1-2.4 keV; Frontera et al. 1998). It is also consistent
with the earlier determination (Galama et al. 1997; see also Reichart
1999).

Next we took a single value for the power-law decay in V, R$_{\rm c}$
and I$_{\rm c}$, computed the sums of these power-law decays plus
supernova light curves, and fitted for the power-law decay index,
$\alpha$, by minimizing the $\chi$-squared of the V, R$_{\rm c}$,
I$_{\rm c}$ light curves and the resulting fit. But first, the
assumption of a constant color that was used to derive a V-band
magnitude from the September 1997 unfiltered STIS observation is no
longer valid as in this hypothesis at late times the SN emission will
dominate the light and so we expect the light to have reddened (see
Fig. \ref{Lightcurve2}). Thus, in order to convert the observed counts
into broadband magnitudes we assumed that the spectral energy
distribution was similar to that of SN\,1998bw, redshifted to $z =
0.695$. Our procedure was as follows: (i) we used the redshifted U, B,
V, R$_{\rm c}$, I$_{\rm c}$ spectral-flux distribution of SN\,1998bw
(on day 111, corresponding to the STIS epoch at 188 days), (ii) we
corrected for Galactic extinction, $A_V = 0.78 \pm 0.12$,
interpolating to the redshifted central wavelengths of the U, B, V,
R$_{\rm c}$, I$_{\rm c}$ bands using the Galactic extinction curve of
Cardelli, Clayton and Mathis (1989) to obtain the spectral flux
distribution as it would be observed in the direction of GRB\,970228,
and (iii) we used this spectral flux distribution in the SYNPHOT
routine CALCPHOT, and applied the Landsman (1997) correction, to
determine corresponding broad-band magnitudes, $V = 28.56$, $R_{\rm c}
= 27.42$ and $I_{\rm c} = 26.43$.  Again, we adopt an uncertainty of
0.4 mag to make some allowance for the sensitivity of the result to
the assumed spectral shape. Finally we minimized the $\chi$-squared
and found $\alpha$ = \gpm{-1.73}{0.09}{0.12} ($\chi^{2}$ = 17.5 with
14 d.o.f; we note that the $\chi$-squared is not strictly correct
since a few of the data points are not independent). The fit is
excellent in view of the uncertainties in the model. Similar results
were presented by Reichart (1999). The fit is somewhat better than
that presented by Reichart (1999), presumably because of our
consistent reanalysis of all the original images, and, in particular,
because our STIS magnitudes were calculated using the knowledge of the
spectral shape of the template supernova.

The model has very few assumptions: the power law index $\alpha$ is
the only free parameter and we did not allow the peak luminosity of
the SN to be a free parameter but used the SN\,1998bw observations as
they are. SN\,1998bw is, of course, an unusual example of the class,
being very bright and rapidly expanding, but its coincidence with
GRB\,980425 makes it the only template we can reasonably choose for
now.  Also, from the `collapsar' model we expect a GRB to be
accompanied by a supernova of type I$_{\rm b/c}$.  In any case, the
light curves (Fig. \ref{Lightcurve2}) as well as the intermediate-time
spectral flux distribution (Fig. \ref{Spectra}) and intermediate-time
colors are remarkably well explained by the sum of a power-law plus
supernova light curve. It can explain the apparent transition from a
rapid to a slow and again a rapid decline of the light curves, is in
good agreement with the upper limits in the light curves and it can
account for the unusual intermediate-time red color $V-I_{\rm c} =
2.24$ and the peaked spectral flux distribution on March 30.8, 1997 UT
(see also Reichart 1999). These features cannot be accounted for by
simple relativistic blast-wave models. We thus find the data to be in
agreement with the hypothesis of Reichart (1999) that a supernova
dominated the light at late times. Together with the evidence for
GRB\,980326 (Bloom et al. 1999) and GRB\,980425 (Galama et al. 1998)
this is further support for the idea that at least some GRBs are
associated with a possibly rare type of supernova.

\acknowledgements We wish to thank Drs. Tuffs, Soifer, Neugebauer,
Guarnieri, Bartolini and Piccioni for observations of GRB\,970228.
This work is partly based on observations collected at the European
Southern Observatory, Paranal, Chile (VLT-UT1 Science Verification
Program). T.J. Galama is supported through a grant by NFRA under
contract 781.76.011. PMV and ER are supported by the NWO Spinoza
grant. CK acknowledges support from NASA grant NAG 5-2560.

{\scriptsize
\begin{table}
\caption[ ]{Summary of optical observations. Upper limits are three
sigma. Abbreviations: OT, optical transient; GAL, host galaxy; RAO,
Rome Astrophysical Observatory; BUT, Bologna University Telescope;
WHT, William Hershell Telescope; APO, Apache Point Observatory; NOT,
Nordic Optical Telescope; INT, Isaac Newton Telescope; CA, Calar Alto
3.5-m telescope; HST, Hubble Space Telescope; P5m, Palomar 200-inch;
VLT-UT1, Very Large Telescope Unit 1. \label{Data1}}
\begin{tabular}{lllll}
Date (UT) & \multicolumn{1}{c}{Telescope} & Exp. time (sec) &
Magnitude& \multicolumn{1}{c}{Remarks} 
\\
\tableline
Feb. 28.81 1997& RAO &    & $R_{\rm c}=20.0 \pm 0.5$ & Unfiltered
OT+GAL \\
Feb. 28.83 1997& BUT &    & $R_{\rm c}=20.58 \pm 0.28$ & OT+GAL\\
Feb. 28.99 1997& WHT &    & $V=21.01 \pm 0.10$ & OT+GAL \\
Feb. 28.99 1997& WHT &    & $I_{\rm c}=20.51 \pm 0.21$& OT+GAL\\
Mar. 03.10 1997& APO &    & $B_{\rm J} = 24.5 \pm 0.7$ & OT+GAL\\
Mar. 03.90 1997& NOT &    & $R_{\rm c}>22.70 \pm 0.20 $ &
Unfiltered OT+GAL\\
Mar. 04.86 1997& NOT &    & $R_{\rm c}=23.28 \pm 0.40$ &  Unfiltered
OT+GAL\\
Mar. 04.87 1997& NOT &    & $V>23.96 \pm 0.12 $ & OT+GAL\\
Mar. 04.89 1997& NOT &    & $B>23.87 \pm 0.10 $ & OT+GAL\\
Mar. 06.32 1997& Keck&    & $R_{\rm c}=23.60 \pm 0.22$ & OT+GAL \\
Mar. 08.86 1997& INT &    & $V>24.00 \pm 0.10 $ & OT+GAL\\
Mar. 08.88 1997& WHT &    & $I_{\rm c}>22.68 \pm 0.10 $ &
OT+GAL\\
Mar. 08.89 1997& INT &    & $R_{\rm c} > 23.35 \pm 0.08 $ &
OT+GAL\\
Mar. 09.85 1997& INT &    & $B=25.43 \pm 0.31$ & OT+GAL\\
Mar. 09.89 1997& INT &    & $R_{\rm c}=23.75 \pm 0.28$ & OT+GAL\\
Mar. 13.00 1997& NTT &    & $R_{\rm c}=24.36 \pm 0.15$ & OT+GAL\\
Mar. 17.84 1997 & CA &   & $J > 21.45 \pm 0.15$ & OT+GAL\\
Mar. 17.83 1997 & CA &   & $H > 20.02 \pm 0.15$ & OT+GAL\\
Mar. 17.81 1997 & CA &   & $K^{'} > 20.13 \pm 0.15$ & OT+GAL\\
Mar. 26.38 1997& HST &    & $V=25.98 \pm 0.22$ & OT\\
Mar. 26.47 1997& HST &    & $I_{\rm c}=23.76 \pm 0.11$ & OT\\
Mar. 30.27 1997 & Keck &    & $K = 22.85 \pm 0.25$ & OT+GAL\\ 
Mar. 30.30 + 31.26 1997 & Keck &    & $J = 23.27 \pm 0.15$ & OT+GAL\\ 
%Mar. 26.38 1997& HST &    & $V=25.8 \pm 0.3$ & GAL\\
%Mar. 26.47 1997& HST &    & $I_{\rm c}=24.6 \pm 0.3$ & GAL\\
Apr. 05.76 1997& Keck&    & $R_{\rm c}=24.97 \pm$ 0.25 & OT+GAL\\ 
Apr. 07.22 1997& HST &    & $V=26.36 \pm 0.26$ & OT\\
Apr. 07.30 1997& HST &    & $I_{\rm c}=24.10 \pm 0.15$ & OT\\
%Apr. 07.22 1997& HST &    & $V=25.7 \pm 0.3$ & GAL\\
%Apr. 07.30 1997& HST &    & $I_{\rm c}=25.8 \pm 0.3$ & GAL\\
Mar. 26 + Apr. 7 1997& HST &    & $V=25.75 \pm 0.35$ & GAL\\
Mar. 26 + Apr. 7 1997& HST &    & $I_{\rm c}=24.7 \pm 0.5$ & GAL\\
Sep. 04    1997& P5m &    & $R=25.5 \pm 0.5$ & GAL\\
Sep. 04.71 1997& HST &    & $V=28.80 \pm 0.40$ & OT\\
Sep. 04.71 1997& HST &    & $R_{\rm c}=27.69 \pm 0.40$ & OT\\
Sep. 04.71 1997& HST &    & $I_{\rm c}=26.54 \pm 0.40$ & OT\\
Sep. 04.71 1997& HST &    & $V=25.77 \pm 0.20$ & GAL\\
Sep. 04.71 1997& HST &    & $R_{\rm c}=25.22 \pm 0.20$ & GAL\\
Sep. 04.71 1997& HST &    & $I_{\rm c}=24.73 \pm 0.20$ & GAL\\
Sep. 01.35 1998& VLT-UT1 & 1800 & $B> 26.08 \pm 0.08 $ & GAL\\
Sep. 01.35 1998& VLT-UT1 & 1200 & $V > 25.33 \pm 0.05 $ & GAL\\
Sep. 01.35 1998& VLT-UT1 & 1500 & $R_{\rm c} > 25.29 \pm 0.06 $
& GAL\\

\tableline
\end{tabular}
\end{table}
}

\begin{table}
\caption[ ]{B, V, R$_{\rm c}$, I$_{\rm c}$ magnitudes, associated
random errors $\Delta$B, $\Delta$V, $\Delta$R$_{\rm c}$,
$\Delta$I$_{\rm c}$, and the coordinates of the reference stars. We
estimate the absolute calibration uncertainty to be 0.05 magnitudes in
B, V, and R$_{\rm c}$ and 0.10 magnitudes in I$_{\rm c}$.
\label{Ref}}
\begin{tabular}{lrlllllllllll}
Star 	& & A & B & C & D & E & F & G & H \\
\tableline
B & &21.693 &19.614 &23.161 &23.654 &22.296 &23.976 \\
$\Delta$B&&0.022&0.010&0.042&0.060&0.016&0.073\\
V &&20.345 &18.793 &21.782 &21.972 &20.648 &22.137 &23.237 &23.003\\
$\Delta$V&&0.011&0.010&0.017&0.021&0.010&0.023&0.064&0.044\\
R$_{\rm c}$& &19.646 &18.413 &20.874 &21.119 &19.640 &20.889 &23.008
&21.906\\
$\Delta$R$_{\rm c}$&&0.040&0.040&0.025&0.032&0.010&0.035&0.103&0.067\\
I$_{\rm c}$ && 18.787 & 17.758 & 20.099 & 20.501 & 18.665 & 19.279 & &
20.923\\
$\Delta$I$_{\rm c}$ && 0.034 & 0.034 & 0.045 & 0.062 & 0.034 & 0.040 &
& 0.080\\
\tableline
%Coordinates (J2000)\\
RA (J2000) &5:01: & 48.9  & 49.3 &  47.8 &  48.7 &  45.9  & 45.6  & 47.4 &
46.4\\
Dec (J2000) & 11: & 45:59 & 45:43 & 46:56 & 47:01 & 47:12 & 47:24 & 46:56 &
46:54\\

\end{tabular}
\end{table}

\end{document}